\documentstyle[aps]{revtex}
\def\a{\alpha}
\def\b{\beta}

\def\e{\epsilon}

\def\f{\phi}
\def\g{\gamma}
\def\h{\eta}

\def\j{\psi}

\def\l{\lambda}

\def\q{\theta}
\def\r{\rho}
\def\s{\sigma}
\def\t{\tau}

\def\x{\xi}

\def\D{\Delta}

\def\J{\Psi}

\def\zb{\bar{z}}

\def\pp {\partial }
\def\pb {\bar{\partial }}
\def\be{\begin{equation}}
\def\ee{\end{equation}}
\def\ben{\begin{eqnarray}}
\def\een{\end{eqnarray}}
\begin{document}
\title{  Painlev\'{e} analysis of the coupled
nonlinear Schr\"{o}dinger equation for polarized optical waves in an isotropic medium }
\author{Q-Han Park\footnote{ Electronic address; qpark@nms.kyunghee.ac.kr }
{and}
H. J. Shin\footnote{ Electronic address; hjshin@nms.kyunghee.ac.kr }}
\address{\begin{center}{\it
Department of Physics \\
and \\
Research Institute of Basic Sciences \\
Kyunghee University,
Seoul, 130-701, Korea}\end{center}}
\maketitle
\def\be{\begin{equation}}
\def\ee{\end{equation}}
\def\ben{\begin{eqnarray}}
\def\een{\end{eqnarray}}
\begin{abstract}
Using the Painlev\'{e} analysis, we investigate the integrability properties of
a system of two coupled nonlinear Schr\"{o}dinger equations that describe
the propagation of orthogonally polarized optical waves in an isotropic medium.
Besides the well-known integrable vector nonlinear Schr\"{o}dinger equation,
we show that there exist a new set of equations passing the Painlev\'{e} test
where the self and cross phase modulational terms are of different magnitude.
We introduce the Hirota bilinearization and the B\"{a}cklund transformation
to obtain soliton solutions and prove integrability by making
a change of variables. The conditions on the third-order susceptibility tensor
$\chi^{(3)} $ imposed by these new integrable equations are explained.
\bigskip
\end{abstract}
\maketitle
\section{Introduction}
The coupling between copropagating optical pulses in a nonlinear medium has led to many
important applications in optical fiber systems such as optical switching and
soliton-dragging logic gates \cite{Islam}. The governing equation for the propagation
of two orthogonally polarized pulses in a monomode birefringent fiber is given by the
coupled nonlinear Schr\"{o}dinger(NLS) equation, where the nonlinear coupling terms
are determined by the third-order susceptibility tensor $\chi^{(3)} $ of the fiber.
In an isotropic medium, the tensor $\chi^{(3)} $ has three independent components
$\chi^{(3)}_{xxyy}, \chi^{(3)}_{xyxy} $ and $ \chi^{(3)}_{xyyx}$
and the nonlinear polarization components which account for the nonlinear coupling terms
take the form
\ben
P_{x}&=& {3\e_{0} \over
2}\Big[ [(\chi^{(3)}_{xxyy}+\chi^{(3)}_{xyxy}+\chi^{(3)}_{xyyx})|E_{x}|^2
+(\chi^{(3)}_{xxyy}+\chi^{(3)}_{xyxy})|E_{y}|^2 ]E_{x}
+\chi^{(3)}_{xyyx}E_{y}^2 E_{x}^{*} \Big], \nonumber \\
P_{y}&=& {3\e_{0} \over
2}\Big[ [ (\chi^{(3)}_{xxyy}+\chi^{(3)}_{xyxy}+\chi^{(3)}_{xyyx})|E_{y}|^2
+(\chi^{(3)}_{xxyy}+\chi^{(3)}_{xyxy})|E_{x}|^2 ]E_{y}
+\chi^{(3)}_{xyyx}E_{x}^2 E_{y}^{*} \Big].
\label{Pnl}
\een
In the case of silicar fibers, $\chi^{(3)}_{xxyy}\approx \chi^{(3)}_{xyxy} \approx
\chi^{(3)}_{xyyx}$ and the nonlinear terms above have a ratio of
3;2;1. However, when the fiber is elliptically birefringent with the ellipticity angle
$\theta \approx 35^{o}$,
and also the beat length due to birefrigence is much smaller than the typical propagation
distances, the coupled NLS equation takes the form of the vector NLS equation whose
nonlinear terms have a ratio of 1;1;0 \cite{Agr},
which is known to be integrable via the inverse scattering method \cite{ZaSha}\cite{zak}.
In general, the coupled NLS equations with arbitrary coefficients are not integrable.
Mathematically, there exists a systematic way of generalizing the NLS equation to the
multi-component cases \cite{fordy} and to the higher-order cases \cite{kps}
using group theory which preserves the integrabiiity structure.
This gives rise to various integrable, coupled NLS equations among $N$ scalar
fields $\j_i ; i = 1,..., N$ with specific set of coupling parameters.
For $N=2$, the vector NLS equation is the only nontrivial integrable equation in the
group theoretic construction.
However, it is not known whether there can be other cases of the integrable coupled NLS
equation for $N=2$ with nonlinear coupling terms as in Eq. (\ref{Pnl}) except for the
vector NLS equation.

In this paper, using the Painlev\'{e} analysis we investigate the integrability properties
of the coupled NLS equation relevant to the propagation of orthogonally polarized
optical waves in an isotropic medium. Motivated by Eq. (\ref{Pnl}), we consider the general
form of the coupled NLS equation such that
\ben
i \pb q_1 &=& \pp ^2 q_1 + q_1 (\gamma  _1 |q_1|^2 + \gamma _2 |q_2|^2 ) + \gamma _3 q_1^*
q_2^2 + \gamma _4 q_1^2 q_2^*, \nonumber \\
i \pb q_2 &=& \b \pp ^2 q_2 + q_2 (\gamma  _2 |q_1|^2 + \gamma _1 |q_2|^2 ) + \gamma _3 q_2^*
q_1^2 + \gamma _4 q_2^2 q_1^* ,
\label{nlseqns}
\een
where $\b = \pm 1$ signify the relative sign of the group-velocity dispersion terms and we
use the notation $\pp = \pp /\pp z , \pb =\pp / \pp \zb $.
We find that the system passes the Painlev\'{e} test whenever the parameters belong to one of
the following four classes;  $(i) ~\beta =1, \gamma_1 = \gamma_2 ,
\gamma_3 = \gamma_4 =0 $,  $(ii) ~\beta =1 , ~ \gamma_2 = 2 \gamma_1 , \gamma_3 = -\gamma_1 ,
\gamma_4 \mbox{ arbitrary }$, $(iii)~ \beta =1, \gamma_2 = 2 \gamma_1 ,
\gamma_3 = \gamma_1 , \gamma_4 = 0$ and $(iv) ~\beta = -1, \gamma_1 =- \gamma_2 ,
\gamma_3 = \gamma_4 =0  $.
Case (i)( and (iv)) is the well-known vector NLS equation. The integrability of cases (i) and (iv) have
been demonstrated by Zakharov and Schulman by deriving an appropriate inverse
scattering formalism \cite{zak}\cite{zak1}.  However, cases (ii) and (iii) are new as far as we know. In particular,
case (ii) corresponds to the propagation in the isotropic nonlinear medium with
the property that $\chi^{(3)}_{xxyy}+\chi^{(3)}_{xyxy}=-2\chi^{(3)}_{xyyx}$.
We find the Hirota bilinearization and the B\"{a}cklund transformation
of cases (ii) and (iii), and compute soliton solutions.  As for the integrability of
cases (ii) and (iii), we prove that they are essentially identical to two independent
NLS equations. This implies that in the case (ii), there is no physical interactions between
two optical pulses with opposite circular polarizations.
We also show that our Painlev\'{e} analysis is consistent with the group theoretical method of
generalizing the integrable NLS equations when the group theoretical method is combined with
the reduction procedure.

\section{Painlev\'{e} analysis of the coupled NLS equation }
The Painlev\'{e} analysis for a partial differential equation was first introduced by
Weiss, Tabor, and Carnevale\cite{weiss} who defined that a partial differential equation
has the Painlev\'{e} property if its general solution is single-valued about the movable
singularity manifold. This method is to seek a solution of a given differential equation
in a series expansion in terms of $\phi (z,\zb ) = z-\j (\zb )$, where $\j (\zb )$ is an
arbitrary
analytic function of $\zb $ and $\phi =0$ defines a non-characteristic movable singularity
manifold. Then, the equation has the Painlev\'{e} property, thus becomes integrable,
if there exists a sufficient number of arbitrary functions in the series solution.
For $\beta =1$, we postulate a solution of the form,
\ben
q_1 &=& \sum_{m \ge 0} R_m (\bar z) (z-\j )^{m-\s}, \nonumber \\
q_1^* &=& \sum_{m \ge 0} S_m (\bar z) (z-\j )^{m-\s}, \nonumber \\
q_2 &=& \sum_{m \ge 0} T_m (\bar z) (z-\j )^{m-\s}, \nonumber \\
q_2^* &=& \sum_{m \ge 0} U_m (\bar z) (z-\j )^{m-\s}.
\label{subst}
\een
Substituting these ans\"{a}tze into Eq. (\ref{nlseqns}) and looking at the leading order behavior,
we find that $\s=1$ and the following equations should be satisfied:
\ben
\gamma _1 U_0^2 T_0 + \gamma _2 R_0 S_0 U_0 + \gamma _3 S_0^2 T_0 + \gamma _4 U_0^2 R_0 + 2 U_0 &=&
0, \nonumber \\
\gamma _1 T_0^2 U_0 + \gamma _2 R_0 S_0 T_0 + \gamma _3 R_0^2 U_0 + \gamma _4 T_0^2 S_0 + 2 T_0 &=&
0, \nonumber \\
\gamma _1 R_0^2 S_0 + \gamma _2 R_0 T_0 U_0 + \gamma _3 T_0^2 S_0 + \gamma _4 R_0^2 U_0 + 2 R_0 &=&
0, \nonumber \\
\gamma _1 S_0^2 R_0 + \gamma _2 S_0 T_0 U_0 + \gamma _3 U_0^2 R_0 + \gamma _4 S_0^2 T_0 + 2 S_0 &=&
0.
\label{init}
\een
In order to facilitate solving Eq. (\ref{init}), we define
$x \equiv U_0 R_0,~ y \equiv T_0 S_0, ~t \equiv R_0 S_0,~ s \equiv
U_0 T_0$ so that the first two equations in Eq. (\ref{init}) can be written as
\ben
\gamma _1 s + \gamma _2 t +2 + \gamma _4 x = -\gamma _3 {ty \over x}, \nonumber \\
\gamma _1 s + \gamma _2 t +2 + \gamma _4 y = -\gamma _3 {tx \over y},
\label{red1}
\een
while the last two as
\ben
\gamma _1 t + \gamma _2 s +2 + \gamma _4 x = -\gamma _3 {sy \over x}, \nonumber \\
\gamma _1 t + \gamma _2 s +2 + \gamma _4 y = -\gamma _3 {sx \over y}.
\label{red2}
\een
Each pair can be combined to give $(x-y)(\gamma _4 - \gamma _3 t {x+y \over xy})=0$,
and $(x-y)(\gamma _4 - \gamma _3 s {x+y \over xy})=0$. One can readily check that solutions
of these equations can be classified in seven different cases,
\vglue .2in
\noindent
(case 1) $x=y, \gamma_{1}=\gamma_{2}+\gamma_{3}$, \\
(case 2) $x=y, t = x $,  \\
(case 3) $x=y, t=-x $,   \\
(case 4) $t=s, \gamma _4 = \gamma _3 t {x+y \over xy}$, \\
(case 5) $x=-y, \gamma_{4}=0, \gamma_{2}=\gamma_{1}+\gamma_{3}, t+s=-2/\gamma_{1} $
\\
(case 6) $\gamma _3=\gamma _4=0, s=t=-2/(\gamma_{1}+\gamma_{2})$.
\\
(case 7) $\gamma_{3}=\gamma_{4}=0, \gamma_{1}=\gamma_{2}, t+s=-2$
\vskip .2in
For each cases, we check the powers, so called resonances, at which the arbitrary functions
can arise in the series solution. Equating coefficients of the $(z-\j)^{j-3}$ term in Eq.
(\ref{nlseqns}) with the ans\"{a}tze in Eq. (\ref{subst}), we obtain a system of four linear
algebraic equations in ($R_j, S_j, T_j, U_j$) which are given in a matrix form by,
\be
Q_{j} \pmatrix{ R_j \cr S_j \cr T_j \cr U_j } =\pmatrix{ F_j \cr G_j \cr H_j \cr K_j }.
\label{recursion}
\ee
The $4 \times 4$ matrix $Q_{j}=(j-1)(j-2) I_{4 \times 4} +
\pmatrix{Q^{(1)}_{j} & Q^{(2)}_{j} \cr
Q^{(3)}_{j} & Q^{(4)}_{j} }$ has block components:
\ben
Q^{(1)}_{j} &=&
\pmatrix{ 2 \gamma _1 R_0 S_0 + \gamma _2 T_0 U_0 +2 \gamma _4 R_0 U_0
&
\gamma _1 R_0^2  + \gamma _3 T_0 ^2 \cr
 \gamma _1 S_0^2  + \gamma _3 U_0 ^2 &
2 \gamma _1 R_0 S_0 + \gamma _2 T_0 U_0 +2 \gamma _4 S_0 T_0    }
,
\nonumber \\
Q^{(2)}_{j} &=&
\pmatrix{\gamma _2 R_0 U_0 + 2 \gamma _3 T_0 S_0
&
\gamma _2 R_0 T_0  + \gamma _4 R_0 ^2 \cr
\gamma _2 S_0 U_0  + \gamma _4 S_0 ^2
&
\gamma _2 S_0 T_0  + 2 \gamma _3 U_0 R_0 } ,
\nonumber \\
Q^{(3)}_{j} &=&
\pmatrix{ \gamma _2 T_0 S_0  + 2 \gamma _3 R_0 U_0
&
\gamma _2 R_0 T_0  + \gamma _4 T_0 ^2
\cr
\gamma _2 U_0 S_0  + \gamma _4 U_0 ^2
&
\gamma _2 R_0 U_0  + 2 \gamma _3 T_0 S_0 } ,
\nonumber \\
Q^{(4)}_{j} &=&
\pmatrix{  2 \gamma _1 T_0 U_0  + \gamma _2 R_0 S_0 +2 \gamma _4 S_0 T_0
&
\gamma _1 T_0^2 + \gamma _3 R_0 ^2
\cr
\gamma _1 U_0^2  +\gamma _3 S_0 ^2
&
2 \gamma _1 T_0 U_0  + \gamma _2 R_0 S_0 +2 \gamma _4 U_0 R_0} ,
\een
and
\ben
F_j &\equiv&
-\sum^{l+m+n=j }_{0 \le l , m, n <j } (\gamma _1 R_l R_m S_n + \gamma _2 R_l T_m U_n
  +\gamma _3 S_l T_m T_n + \gamma _4 R_l R_m U_n) + i R_{j-2} ^{'} - i (j-2) \j^{'} R_{j-
1} , \nonumber \\
G_j &\equiv&
-\sum^{l+m+n=j }_{0 \le l , m, n <j } (\gamma _1 S_l S_m R_n + \gamma _2 S_l U_m T_n
  +\gamma _3 R_l U_m U_n + \gamma _4 S_l S_m T_n) - i S_{j-2} ^{'} + i (j-2) \j^{'} S_{j-
1} , \nonumber \\
H_j &\equiv&
-\sum^{l+m+n=j }_{0 \le l , m, n <j } (\gamma _1 T_l T_m U_n + \gamma _2 T_l R_m S_n
  +\gamma _3 U_l R_m R_n + \gamma _4 T_l T_m S_n) + i T_{j-2} ^{'} - i (j-2) \j^{'} T_{j-
1} , \nonumber \\
K_j &\equiv&
-\sum^{l+m+n=j }_{0 \le l , m, n <j } (\gamma _1 U_l U_m T_n + \gamma _2 U_l S_m R_n
  +\gamma _3 T_l S_m S_n + \gamma _4 U_l U_m R_n) - i U_{j-2} ^{'} +i (j-2) \j^{'} U_{j-
1} .
\een
The resonances occur when $\det Q_{j} = 0$. Now, we compute the resonance values and check
the Painlev\'{e} property of Eq. (\ref{nlseqns}) for each seven cases as introduced above.
\vskip .2in
\begin{center}
{\bf Case 1. $\bf x=y; ~ \gamma_{1}=\gamma_{2}+\gamma_{3}$}
\end{center}
\vskip .2in
In this case, we can solve for $T_{0} , R_{0}$  such that
\be
T_0={-2 U_0 \over \gamma _1 (S_0^2 + U_0^2) +\gamma _4 S_0 U_0}, \ \
R_0={-2 S_0 \over \gamma _1 (S_0^2 + U_0^2) +\gamma _4 S_0 U_0} .
\label{case1a1}
\ee
When we substitute these solutions into the resonance condition, $\det Q_{j} = 0$, we find that
the resonances do not occur at the integer values of $j$.  Therefore, this case does not pass
the Painlev\'{e} test for integrability.
\vskip .2in
\begin{center}
{\bf Case 2 and Case 3. $\bf x=y;~~ t=\pm x$ }
\end{center}
\vskip .2in
We have  solutions,
\be
S_0=\pm U_0={-2 \over \gamma _1 + \gamma _2 +\gamma _3 \pm \gamma _4} {1 \over R_0},\ \ T_0= \pm R_0,
\ee
where $+$ and $-$ sign correspond to the Case 2 and the Case 3 respectively.
Substituting these solutions into the resonance condition $\det Q_{j} = 0$, we
find that the resonance values  $j=-1, 0, 1, 1, 2, 2, 3, 4$ occur when $\gamma _2=2 \gamma _1,
\gamma _3= \pm \gamma _4+\gamma _1$. The resonance $j=-1$ is related with the arbitrariness of
$\j$, while the resonance $j=0$ is related with the arbitrariness of $R_0$. The recursion
relation in Eq. (\ref{recursion}) determines $R_1, S_1, T_1, U_1$ in terms of
$R_0, S_0, T_0, U_0, \j$. The degree of multiplicity of the resonance $j=1$ is two
and it turns out that there exist two arbitrary functions consistently only if $\gamma_{4}=0$.
Therefore, the case where $\gamma_{2} = 2 \gamma_{1}, \gamma_{3} = \gamma_{1} $
and $\gamma_{4}=0$ passes the Painlev\'{e} test.
\vskip .2in
\begin{center}
{\bf Case 4. $\bf t=s; ~ \gamma_{4}xy=\gamma_{3}t(x+y) $}
\end{center}
\vskip .2in
Eq. (\ref{init}) together with the condition $t=s, ~ \gamma_{4}xy=\gamma_{3}t(x+y) $
results in
\be
t = {-2 \over \gamma _1 + \gamma _2 -\gamma _3 + (\gamma _4 ^2 / \gamma _3) },\ \
x = ({\gamma _4 \over 2 \gamma _3 } \pm \sqrt{(\gamma _4/2 \gamma _3)^2 -1}) t ,
\ee
and
\be
S_0 = {t^2 \over x} {1 \over T_0},\ \ U_0 = {t \over T_0},\ \ R_0 = {x \over t}
T_0.
\label{zerosol}
\ee
When we substitute these solutions into the resonance condition $\det Q_{j} = 0$, we obtain
\be
(j-4)(j-3)j(j+1)(j^2-3j+2 {\gamma _4 ^2 -4 \gamma _3 ^2 \over \gamma _4 ^2 -\gamma _3 ^2 + \gamma _2 \gamma _3
+ \gamma _3 \gamma_1 })(j^2-3j+2 {\gamma _4 ^2 -2 \gamma _3 ^2 +2 \gamma _2 \gamma_3 -2\gamma_1
\gamma _3 \over \gamma _4 ^2 -\gamma _3 ^2 + \gamma _2 \gamma _3 + \gamma_1 \gamma _3})=0.
\label{deter}
\ee
Note that the Painlev\'{e} test requires the resonances $j$ to be integers and
the degeneracy of resonance at $j=0$ to be one since there is only one arbitrary function
$T_0$ as in Eq. (\ref{zerosol}).  This requirement leads to the result,
$ \gamma _2=2 \gamma_{1} , \gamma _3= -\gamma_{1}  $ and $\gamma_{4}$ arbitrary, so that
resonances are $j=-1,0,1,1,2,2,3,4$. The
recursion relation in Eq. (\ref{recursion}) determines $T_1, U_1, T_2, U_2$ such as
\ben
T_1 &=& {1 \over 4} (\sqrt{\gamma _4 ^2 -4}-\gamma _4 )(2 R_1 + i\sqrt{\gamma _4 ^2 -4} T_0
\j_x ), \nonumber \\
U_1 &=& -{1 \over 2} (\sqrt{\gamma _4 ^2 -4}+\gamma _4 )( S_1 + {i\over \sqrt{\gamma _4 ^2 -4}}
{\j_x \over T_0} ), \nonumber \\
T_2 &=& {1 \over 2} (\sqrt{\gamma _4 ^2 -4}-\gamma _4 )(R_2 + {\sqrt{\gamma _4 ^2 -4} \over 12}
( T_0 \j_x ^2 + 2 i{\pp T_0 \over \pp x} )  ), \nonumber \\
U_2 &=& {1 \over 12} (\sqrt{\gamma _4 ^2 -4}+\gamma _4 )(-6 S_2 + {i\over \sqrt{\gamma _4 ^2 -
4}} ({\j_x ^2 \over T_0} +2 i{1 \over T_0 ^2 } {\pp T_0 \over \pp x})).
\een
Similarly, $R_3, T_3, U_3$ are determined in terms of $\j, T_0, R_1, S_1, R_2, S_2$.
In the same way, we can check that there exists one arbitrary function at the j=4
resonance and no more arbitrary functions in higher lebels. All these facts have been
confirmed with the symbolic manipulation program Macsyma.
Thus, the system passes the Painlev\'{e} test
when  $ \gamma _2=2 \gamma_{1} , \gamma _3= -\gamma_{1}  $ and $\gamma_{4}$ arbitrary.
We show that this case is indeed integrable in Sec. 3.
\vskip .2in
\begin{center}
{\bf Case 5. $\bf x=-y; ~ \gamma_{4}=0; \gamma_{2}=\gamma_{1}+\gamma_{3}; t+s=-2/\gamma_{1} $}
\end{center}
\vskip .2in
In this case, the resonances are
at $j=-1, 0, 0, 3, 3, 4, {3 \over 2} \pm \sqrt{9 +16 \gamma _3
/ \gamma _1 }$, which in turn requires that $\gamma _1 = -2 \gamma _3,\ \ \gamma _2 = -\gamma _3.$ But
inconsistency among the four equations in Eq. (\ref{recursion}) arises at the $j=2$
level, so that the Painlev\'{e} test fails.
\begin{center}
{\bf Case 6. $\bf \gamma_{3}=\gamma_{4}=0;  s=t=-2/(\gamma_{1}+\gamma_{2})$}
\end{center}
\vskip .2in
The resonance condition, $det Q_{j}=0$, leads to the following solutions;
\be
j = -1, 0, 0, 3, 3, 4, {3 \over 2} \pm {1 \over 2(\gamma _1 + \gamma _2)} \sqrt{25 \gamma _1 ^2
+ 18 \gamma _1 \gamma _2 -7 \gamma _2 ^2}.
\ee
The integer resonances occur if (i) $\gamma _2 = 3 \gamma _1$, or (ii) $\gamma _2 = - \gamma _1$.
The first case (i) leads to inconsistencies among four equations in Eq. (\ref{recursion})
at j=2, while the second case (ii) similarly leads to
inconsistency at $j=0$. Therefore,  the Painlev\'{e} test fails in this case.
\vskip .2in
\begin{center}
{\bf Case 7. $\bf \gamma_{3}=\gamma_{4}=0; ~ \gamma_{1}=\gamma_{2} ; ~ t+s = -2 $ }
\end{center}
\vskip .2in
This case corresponds to the well-known integrable vector NLS equation considered by
Zakharov and Schulman \cite{zak}. Together with the parameters;
$\gamma _1 = \gamma _2, \gamma _3 = \gamma _4 =0$, Eq. (\ref{init}) reduces to
\be
2 + \gamma _1 ( T_0 U_0 + R_0 S_0 ) = 0.
\ee
The resonances are  $j=-1, 0, 0, 0, 3, 3, 3, 4$, and it has been checked that
the proper number of arbitrary functions exist. Thus, this case passes the Painlev\'{e} test.
\vskip .2in

So far, we have considered the case where $\beta =1$ in Eq. (\ref{nlseqns}).
For $\beta = -1$,
using the notion of the degenerate dispersion law, Zahkarov and Schulmann found another
integrable theory with anomalous dispersive term \cite{zak1}.
The Painlev\'{e} analysis for the $\beta = -1$ case can be done in the same way as for the
$\beta =1$ case. Thus, we suppress the details of analysis and simply state the results.
The leading order equation is given by
\ben
\gamma _1 s + \gamma _2 t -2 + \gamma _4 x = -\gamma _3 {ty \over x}, \nonumber \\
\gamma _1 s + \gamma _2 t -2 + \gamma _4 y = -\gamma _3 {tx \over y}, \nonumber \\
\gamma _1 t + \gamma _2 s +2 + \gamma _4 x = -\gamma _3 {sy \over x} \nonumber \\
\gamma _1 t + \gamma _2 s +2 + \gamma _4 y = -\gamma _3 {sx \over y},
\label{red3}
\een
whose solutions can be grouped into five distinct cases;
\vglue .2in
\noindent
(case 1) $x=y$ \\
(case 2) $\gamma _4 =0, x=-y, \gamma _3=\gamma _1 +\gamma _2$  \\
(case 3) $\gamma _4 =0, x=-y, t=-s $ \\
(case 4) $\gamma _3 = \gamma _4 =0, \gamma_{1} = -\gamma_{2} $ \\
(case 5) $\gamma _3 = \gamma _4 =0, t =-s$   .
\vskip .2in
Here, only the case 4 passes the Painlev\'{e} test. In this case,
$S = (T_0 U_0 -2 )/ R_0 $ and resonances are $j=-1, 0, 0, 0, 3,
3, 3, 4$. This is the integrable system found by Zakharov and Schulmann \cite{zak}.
All other cases lead to inconsistencies at $j=1$ level thus failing
the Painlev\'{e} test.

\section{Hirota bilinearization and solitons}
One of the main result of the Painlev\'{e} test is to find a new case of coupled
NLS equation in Eq. (\ref{nlseqns}) with parameters given by
$\gamma_2 =2 \gamma_1 , \gamma_3 = - \gamma_1 $ and
$\gamma_4 $ arbitrary. With an appropriate scaling, we can always set the nonzero
$\gamma_1 $ to one. Also, as we show in Sec. 4, we can set $\gamma_4 $ to zero.
From now on, we restrict
to this case ($\beta =1, \gamma_1 =1, \gamma_2 =2, \gamma_3 = -1, \gamma_4 =0$)
and analyze its solution and integrability structures.
It is well known that the Painlev\'{e} analysis in the preceding section can be related
to the B\"{a}cklund transformation (BT). In order to derive the BT, we truncate the
series in Eq. (\ref{subst}) up to a constant level term and substitute $(z-\j)$ by
an arbitrary function $\f (z, \zb)$ to be determined later. Then, the corresponding BT
is given by
\be
q_1 = {R_0 \over \f} + R_1,\ \ q_2 = {T_0 \over \f} + T_1,
\label{BT}
\ee
where the set $(R_1,T_1)$ is a known solution of the coupled NLS equations,
which we assume to be the trivial solution $R_1=T_1=0$. In order for the new set $(q_1 , q_2 )$
to be also a solution, the following equations should hold  \cite{sa}
\ben
i \f \bar{D} R_0 \cdot \f = i \f D ^2 R_0 \cdot \f - R_0 D ^2 \f \cdot \f + R_0^2
R_0^* +2 R_0 T_0 T_0^* - R_0^* T_0^2 \nonumber \\
i \f \bar{D} T_0 \cdot \f = i \f D ^2 T_0 \cdot \f - T_0 D ^2 \f \cdot \f + T_0^2
T_0^* +2 T_0 R_0 R_0^* - T_0^* R_0^2, \nonumber \\
\label{hirota}
\een
Here, the Hirota's bilinears $D$ and $\bar{D}$ are defined by
\be
\bar{D} ^n D ^m f \cdot g = \Big({\pp \over \pp \zb }-{\pp \over \pp \zb ' }\Big) ^n
\Big({\pp \over \pp z}-{\pp \over \pp z'}\Big) ^m
f(z,\zb ) g(z^{'},\zb^{'})\Big|_{z=z^{'} \atop \zb =\zb^{'}}.
\ee
Equation (\ref{hirota}) can be decoupled as
\ben
R_0 D ^2 \f \cdot \f -( \gamma _1 R_0^2 R_0^* + \gamma _2 R_0 T_0 T_0^* + \gamma _3 R_0^* T_0^2 ) &=& \l_1 R_0
\f \cdot \f, \nonumber \\
T_0 D ^2 \f \cdot \f -( \gamma _1 T_0^2 T_0^* + \gamma _2 T_0 R_0 R_0^* + \gamma _3 T_0^* R_0^2 ) &=& \l_2 T_0
\f \cdot \f, \nonumber \\
i \bar{D} R_0 \cdot \f &=& D ^2 R_0 \cdot \f - \l_1 R_0 \cdot \f , \nonumber \\
i \bar{D} T_0 \cdot \f &=& D ^2 T_0 \cdot \f - \l_2 T_0 \cdot \f.
\label{HIR}
\een
Now, explicit N-solitons can be constructed in the usual way by solving $\f, R_0, T_0$
in terms of power series.
\subsection{one-soliton}
For one soliton solution, we choose $\l_{1}=\l_{2}=0$ and assume solutions in a series
form in $\e $ such that $\f =1 + \e^2 h, ~ R_0=\e R, ~ T_0 = \e T$.
Then, by equating the coefficients of the polynomials to zero in Eq. (\ref{HIR}) and solving
them explicitly, we obtain
\be
R = \a \exp\Big( i (a^2-b^2) \zb + 2ab\zb + i a z + b z \Big) ,\ \ T =\b \exp\Big(
i (a^2-b^2) \zb + 2ab \zb +i a z + b z \Big),
\ee
where $\a, \b$ are arbitrary complex numbers while $a,b$ are arbitrary real numbers.
$h$ is also obtained by solving the third order equation such that
\be
h={1 \over 8 b^2} (|\a|^2 + 2 |\b|^2 - {\a^* \b^2 \over \a}) \exp(2 b z+ 4 ab \zb
).
\ee
Consistency requires that phases of the complex numbers $\a$ and $ \b$
should be either the same, or differ by $\pi /2$.
In the case of the same phase, we parameterize $\a $ and $\b$ by
\be
\a = \sqrt{8} b \cos k e^{\D +i \q},\ \ \b = \sqrt{8} b \sin k e^{\D +i
\q},
\ee
in terms of arbitrary real numbers $k, \q, \D$.
Then, the final form of the one soliton solution is given by
substituting $\e =1$ in Eq. (\ref{BT}) such that
\ben
q_1 &=& \sqrt{2} b \cos k  e^{i (a^2-b^2)\zb + i a z +i \q} {\rm sech}
(bz+2ab\zb +\D), \nonumber \\
q_2 &=& \sqrt{2} b \sin k  e^{i (a^2-b^2)\zb + i a z+i \q} {\rm sech}
(bz+2ab\zb+\D).
\een
In the case where phases differ by $\pi /2$, $\a $ and $\b$ are given by
\be
\b = \pm i \a = \pm i \sqrt{8} b  e^{\D +i \q}.
\ee
Then, the corresponding one-soliton solution is
\ben
q_1 &=& \sqrt{2} b  e^{i (a^2-b^2)\zb + i a z +i \q}
{\rm sech} (b z+2ab \zb+\D) , \nonumber \\
q_2 &=& \pm i \sqrt{2} b e^{i (a^2-b^2)\zb + i a z +i \q} {\rm sech}
(b z+2ab \zb+\D).
\een

\subsection{two-soliton}
The two-soliton solution can be obtained using the series expansion $\f=1 + \e^2
h_1 + \e^4 h_2,\ \ R_0=\e \r_1 + \e^3 \r_2,\ \ T_0=\e \t_1 + \e^3 \t_2$.
Inserting these ans\"{a}tze into Eq. (\ref{HIR}), we obtain solutions
\ben
\r_1 &=& f+g , ~ \t_1 = i \r_1 ; ~~
f \equiv e^{-i k^2 \zb + k z +\h_f} , ~ g \equiv  e^{-i l^2 \zb + l z +\h_g
},
\nonumber \\
h_1 &=& 2 \left({f f^* \over (k+k^* )^2} + {f g^* \over (k+l^* )^2} + {g f^* \over
(l+k^* )^2} + {g g^* \over (l+l^* )^2} \right),
\een
where $k, l, \h_f, \h_g$ are arbitrary complex numbers.
Also, after a lengthy but straightforward calculation we obtain
\ben
\r_2 &=& 2 (l-k)^2 \left( {f f^* g \over (k+k^*)^2 (l+k^*)^2} + {f g g^* \over
(k+l^*)^2 (l+l^*)^2} \right)
, ~~
\t_2 = i \r_2, \nonumber \\
h_2 &=& {4 (l-k)^2 (l^*-k^*)^2 f f^* g g^* \over (k+k^*)^2 (l+k^*)^2 (k+l^*)^2
(l+l^*)^2}.
\een
Finally, the two-soliton solution is obtained by taking $\e =1$ in
the BT equation $q_1 = {R_0 \over \f}, q_2 = { T_0 \over \f } $.

Surprisingly, there exists a different type two-soliton solution which can be
obtained by a simple linear superposition of the left-polarized one-soliton with
the right-polarized one-soliton;
\be
q_1 = {f \over \f_1} + {g \over \f_2}, \ \ q_2 = i{f \over \f_1} -i {g \over \f_2},
\ee
where $\f_1 = 1 + 2 f f^* / (k+k^* )^2, \f_2 = 1 + 2 g g^* / (l+l^* )^2  $.
The reason underlying the existence of such a linear superposition
is explained in the following section.

\section{Integrability }
The Painlev\'{e} test in Sec. 2 suggests new integrable cases of coupled NLS equations.
As we have shown in the preceding section, the coupled NLS equation with
$\g_1 =1, \g_2 =2 , \g_3 = -1, \g_4 = 0 $ possesses exact soliton solutions, which reflects
the integrability of the equation. Before proving the integrability by deriving the
corresponding Lax pair, we first note that taking $\g_4 =0$ is not essential.
Make a change of variables such that
\be
Q_1 = x q_1 + y q_2,~~  Q_2 =  y q_1 + x q_2.
\label{cil}
\ee
If $(Q_1 , Q_2 )$ satisfy the coupled NLS equation in Eq. (\ref{nlseqns}) with
$\g_1 =1, \g_2 =2 , \g_3 = -1, \g_4 = 0 $ ,
then $(q_1 , q_2 )$ satisfy Eq. (\ref{nlseqns}) but with parameters
$\g_1 =1, \g_2 =2 , \g_3 = -1, \gamma _4={4xy \over x^2+y^2}$. Thus, we set $\g_4 $ to zero
without loss of generality.
The integrability and the Lax pair of the coupled NLS equation in Eq. (\ref{nlseqns}) with
$\g_1 =1, \g_2 =2 , \g_3 = \pm 1 , \g_4 =0$ follows from the observation that these equations
can be embeded in the integrable coupled NLS equation based on the symmetric space $Sp(2)/U(2)$
given by\footnote{The generalization of NLS equation using symmetric spaces and the concept of
matrix potential can be found in \cite{fordy,kps,bps,park1,park2}}
\ben
i \pb \j_{1} &=&  \Big[  \pp^2 \psi_{1}+ 2 \psi
_{1}^{2}\psi^{*}_{1}+4 \psi_{1}\psi_{2}\psi^{*}_{2}
+2 \psi_{2}^{2}\psi^{*}_{3} \Big],
\nonumber \\
i \pb \j_{2} &=&   \Big[  \pp^2 \psi_{2}+2 \psi_{2}\psi_{1}\psi^{*}_{1}
+2 \psi_{2}^{2} \psi^{*}_{2}+2 \psi_{3}\psi_{1}\psi^{*}_{2}+
2 \psi_{3}\psi_{2}\psi^{*}_{3}
 \Big],
\nonumber \\
i \pb \j_{3} &=&  \Big[  \pp^2 \psi_{3}+2 \psi_{3}^{2}\psi^{*}_{3}+4 \psi_{3}\psi_{2}\psi^{*}_{2}
+2 \psi_{2}^{2}\psi^{*}_{1} \Big].
\label{sp}
\een
Consistent reductions can be made if we take $\j_{1} = \pm \j_{3} $, which are precisely the
cases $\g_1 =2, \g_2 =4 , \g_3 = \pm 2, \g_4 = 0 $ in Eq. (\ref{nlseqns}).  Furthermore,
Eq. (\ref{sp}) arises from the Lax pair
\be
 L_{z} \Psi \equiv  \Big[ \pp + E + \l T \Big] \Psi =0, ~~
 L_{\bar{z}} \Psi \equiv \Big[ \pb + ({1 \over 2}[ E, ~ \tilde{E}  ] - \pp \tilde{E}  )
  - \l E - \l^2 T  \Big] \Psi =0
\label{Lax1}
\ee
where the $4\times 4$ matrices $E$ and $T$ are
\be
E = \pmatrix{0 & 0 &  \j_{1} & \j_{2} \cr
            0 & 0 &  \j_{2} & \j_{3} \cr
            -\j_{1}^{*} & -\j_{2}^{*} & 0 & 0 \cr
            -\j_{2}^{*} & -\j_{3}^{*} & 0 & 0 } , ~
T = \pmatrix{i/2 & 0 & 0 & 0 \cr
            0 & i/2 & 0 & 0 \cr
            0 & 0 & -i/2 & 0 \cr
            0 & 0 & 0 & -i/2 }
\label{emat}
\ee
By taking $\j_{1} = \pm \j_{3}$ in Eqs. (\ref{Lax1}) and (\ref{emat}), we obtain the
Lax pair for the coupled NLS equation in Eq. (\ref{nlseqns}) with
$\g_1 =2, \g_2 =4 , \g_3 = \pm 2 , \g_4 =0$.

More directly, the integrability can be shown by mapping the coupled NLS equation into two
independent (decoupled) NLS equations as follows;
if we substitute
\be
\J_1 = q_1 + i q_2, \ \ \J_2= q_1 - i q_2,
\label{lt}
\ee
in the two independent NLS equations, $i \pb {\J_k} = \pp^2 \J_k +2 |\J_k|^2
\J_k ; k=1,2$, we recover Eq. (\ref{nlseqns}) with
$\gamma _1=2, \gamma _2=4, \gamma _3=-2, \gamma _4=0$. Similarly
using the substitution $\J_1 = q_1 + q_2, \J_2= q_1 - q_2$, we
obtain Eq. (\ref{nlseqns}) with $\gamma _1=2, \gamma _2=4, \gamma _3=2, \gamma _4=0$.
This explains why the linear superposition of two solitons was possible in the previous
section. The decomposition of the coupled NLS equation into two independent NLS equations
implies that the linear combination of solutions according to Eq. (\ref{lt}) becomes a solution
of the coupled NLS equation. Group theoretically, such a decomposition corresponds to
the embedding of symmetric spaces, $(SU(2)/U(1)) \times (SU(2)/U(1)) \subset Sp(2)/U(2)$.
According to the group theoretic construction of the NLS equation using Hermitian symmetric
spaces \cite{helgason}, the above embedding results in two decoupled NLS equations.
It is interesting to see that this decoupling behavior is also reflected in the Painlev\'{e}
analysis. Besides the solution of the leading order equation (\ref{init}) (the case 4 in Sec. 2)
which enables the present coupled NLS equation to pass the Painlev\'{e} test,
for the set of parameters $\g_2 = 2 \g_1 , \g_3 = -\g_1 $, we have another set of solutions of
the leading order equation (\ref{init}),
\be
U_0 = {-2 T_0 \over T_0^2 + R_0^2},\ \ S_0 ={-2 R_0 \over T_0^2 +
R_0^2}.
\ee
This has resonances at $j=-1, -1, 0, 0, 3, 3, 4, 4$. This solution also passes the test.
Note that all resonances are double poles and each poles are precisely those of the NLS equation.
This suggests that the systems under consideration are indeed two independent NLS systems.

So far, we have restricted to the case $\b =1$. For $\b =-1$, our Painlev\'{e} analysis showed
that the only integrable case is the vector NLS equation considered by
Zakharov and Schulmann,
\be
i \pb \J = \pp^2 \J + \J \x \J,\ \ \
-i \pb \x = \pp^2 \x + \x \J \x,
\ee
where $\J =(\j_1, \j_2) $ and $ \x =(\chi_1, \chi_2)$.
Using the reduction $\x = \J^* A$ with $A= \pmatrix{-1 & 0 \cr 0 & 1 }$
and substituting $q_1 = \j_1, q_2 =\j_2 ^*$, one can recover the vector
NLS equation as in Eq. (\ref{nlseqns}) with $\b =-1, \g_1 = -\g_2 , \g_3 =\g_4 =0$.

In a similar vein, we construct a new integrable equation with $\b = -1$
which resembles the previous decoupling NLS equation with $\b =1$.
We take
\be
M = \pmatrix{ \chi_1 & \chi_2 \cr \chi_2 & -\chi_1 },
~~N = \pmatrix{ -\chi_1^* & \chi_2 ^* \cr \chi_2 ^* & \chi_1 ^*},
\label{zs1}
\ee
and define the coupled NLS equation by
\ben
i \pb M &=& \pp^2 M -2MNM, \nonumber \\
-i \pb N &=& \pp^2 N - 2NMN.
\label{zs2}
\een
We find that Eq. (\ref{zs2}) arises from the Lax pair ($[L_z, L_{\bar z}]=0$)
\ben
L_z &=& \pp + \pmatrix{0 & M \cr N & 0} +i {\l \over 2}
\pmatrix{I_{2 \times 2} & 0 \cr 0 & -I_{2 \times 2}}, \nonumber \\
L_{\bar z} &=& \pb -i \pmatrix{0 & \pp M \cr -\pp N & 0} -i \pmatrix{MN & 0 \cr 0 & -NM}
-\l \pmatrix{0 & M \cr N & 0} -{i \over 2} \l^2
\pmatrix{I_{2 \times 2} & 0 \cr 0 & -I_{2 \times 2}}.
\een
If we substitute $q_1 = \chi_1, q_2 = \chi_2 ^*$, we have an integrable
equation with anomalous dispersion term and asymmetric coupling,
\ben
i \pb q_1 &=& \pp^2 q_1 + 2( |q_1|^2 q_1 -2 |q_2|^2 q_1 -q_2^{*2} q_1^*), \nonumber \\
i \pb q_2 &=& -\pp^2 q_2 + 2( |q_2|^2 q_2 -2 |q_1|^2 q_2 -q_1^{*2} q_2^*).
\een
This equation does not belong to the coupled NLS equation in Eq. (\ref{nlseqns})
which has been Painlev\'{e} tested.
\section{Discussion}
In this paper, we have performed a Painlev\'{e} analysis for coupled NLS
equations with coherent coupling terms as given in Eq. (\ref{nlseqns}).
Besides the well-known vector NLS equation ($\b=\pm 1; \g_2 =\pm \g_1 , \g_3 = \g_4 =0)$,
we have found  new integrable cases which are defined by the set of parameters
with $\b =1$; (i)~ $\g_2 = 2 \g_1 , \g_3 = -\g_1 , \g_4 $ arbitrary ,
or (ii) ~ $\g_2 = 2 \g_1 , \g_3 = \g_1 , \g_4 =0$. Painlev\'{e} analysis
shows that these are the only integrable cases except the vector
NLS equation. We have shown that these new equations are essentially identical
to two independent sets of NLS equations. Physically, the first case describes the propagation of optical pulses in
an isotropic nonlinear medium in which the third-order susceptibility tensor satisfies that
$\chi^{(3)}_{xxyy}+\chi^{(3)}_{xyxy}=-2\chi^{(3)}_{xyyx}$, while the second case
does not have a similar interpretation.
The linear transformation in Eq. (\ref{lt}), which decouples the interacting
NLS equation (case (i)) into two independent NLS equations,  also maps two
orthogonal linearly polarized lights into the left and the right circularly polarized lights.
Thus, in such an isotropic medium, left and right circularly polarized lights
do not interact each other thereby preserving circular polarizations.
This case may be compared with a polarization preserving fiber where only one
particular polarization direction is preserved.
It would be interesting to know whether there exists nonlinear isotropic
materials possessing this property.

\vglue .3in
{\bf ACKNOWLEDGEMENT}
\vglue .2in
This work was supported in part by Korea Science and Engineering Foundation
(KOSEF) 971-0201-004-2/97-07-02-02-01-3, and by the program of Basic Science Research, Ministry
of Education BSRI-97-2442, and by KOSEF through CTP/SNU.
\vglue .3in

\end{document}